\documentstyle[aps,amsfonts,twocolumn,epsfig]{revtex}

\begin{document}
\draft

\title{Atomic wave packet basis for quantum information}
\author{Ashok Muthukrishnan \cite{email} and C. R. Stroud, Jr.}
\address{The Institute of Optics, University of Rochester,
Rochester, New York 14627}
\date{March 15, 2001}
\maketitle

%
\newcommand{\hs}[1]{\hspace{#1 ex}}
\newcommand{\vs}[1]{\vspace{#1 ex}}

\newcommand{\ket}[1]{\mbox{$|#1\rangle$}}
\newcommand{\bra}[1]{\mbox{$\langle #1|$}}
\newcommand{\braket}[2]{\mbox{$\langle #1|#2 \rangle$}}
\newcommand{\ketbra}[2]{\mbox{$|#1\rangle\langle #2|$}}
\newcommand{\matrixelem}[3]{\mbox{$\langle #1|#2|#3 \rangle$}}

\newcommand{\cA}{\mbox{${\cal A}$}}
\newcommand{\cB}{\mbox{${\cal B}$}}
\newcommand{\cO}{\mbox{${\cal O}$}}

\newcommand{\nbar}{\bar{n}}

\newcommand{\Sch}{Schr\"odinger }

\newcommand{\postscript}[1] {\centerline{\epsfbox{#1}}}
%

\widetext \hskip.5in
\begin{minipage}{5.75in}
\begin{abstract}
\vs{-7}

\hs{1.5} We propose a wave packet basis for storing and processing several
qubits of quantum information in a single multilevel atom. Using radially
localized wave packet states in the Rydberg atom, we construct an orthogonal
basis that is related to the usual energy level basis by a quantum Fourier
transform. A transform-limited laser pulse that is short compared with the
classical Kepler period of the system interacts mainly with the wave packet
state localized near the atomic core, allowing selective control in this basis.
We argue that wave packet control in this regime is useful for multilevel
quantum information processing.


\end{abstract}
\end{minipage}

\narrowtext \vs{4.2}

%
Quantum information processing is a growing interdisciplinary field that has
seen rapid progress in the last few years \cite{Bouwmeester00}. However
physical implementations of large scale quantum networks are severely limited
by decoherence \cite{Unruh95}, the loss of coherence in the superposition state
of the system. This is especially true of extended systems that involve
entanglement of macroscopic degrees of freedom, such as the center-of-mass
motion of atoms in a trap \cite{Sackett00}.

We consider multiple computational levels in each atom to store information
more densely. The radiative lifetime of Rydberg levels scales as $n^3$, where
$n$ is the principal quantum number, and approaches a millisecond for $n
> 100$. In most atomic quantum logic schemes, the bottleneck for the coherence
time is the macroscopic entanglement of two or more atoms. The multilevel
approach minimizes this decoherence by reducing the number of atoms in the
quantum network, with each $d$-level atom representing $\mbox{log}_2 d$ qubits
of information \cite{Muthukrishnan00}. In a $d=8$ level scheme, for example,
the number of atoms that need to be entangled is less by a factor of three.

A great deal of research has been carried out in the past few years developing
techniques for controlling multilevel Rydberg wave packets using shaped short
pulses.  It is attractive to exploit  this work by developing an alternative
wave packet basis for the quantum information stored in the atomic energy
levels. Rydberg wave packets are coherent superpositions of atomic energy
eigenstates with large principal quantum numbers. We can use the relative
phases of the different eigenstates in the superposition to store information.
Using the energy-time Fourier kernel in unitary time evolution, we define an
orthogonal, discrete wave packet basis that is related to the energy levels by
a quantum Fourier transform. This transform plays a key role in quantum
computing applications \cite{Jozsa98}, and we show how it naturally permits
complementary bases for representing the information stored in a multilevel
quantum system.

In the optical domain, a discrete Fourier transform connects the spectral and
temporal modes of the radiation field in a cavity. Iaconis et al.
\cite{Iaconis00} use this \linebreak

\begin{minipage}{58ex}\vs{24}
\end{minipage}

\noindent approach to characterize the Husimi distribution of a multimode
quantum field in heterodyne detection. Tittel et al. \cite{Tittel00} exploit
the complementarity of energy and time in their scheme for quantum cryptography
using entangled photons. In the atomic case, Muller and Noordam \cite{Muller99}
introduce a Fourier relation between the energy quantum number and a continuous
phase coordinate to study the photoexcitation of Rydberg states. We show how
the quantum Fourier transform of the energy levels corresponds exactly to a
discrete temporal basis of wave packets in cyclic motion about the nucleus.

We consider radial wave packets, which are superpositions of Rydberg states
with low angular momentum. These states are radially localized in probability
and evolve periodically for short times, or in the classical correspondence
limit \cite{Parker86}. The wave packet basis we introduce corresponds to
discrete times in this classical evolution, when each wave packet state is
centered at a different radius from the atomic nucleus. This basis is
non-stationary in time, with wave function amplitudes evolving cyclically
during a classical period. Due to the unequal spacing of the energy levels, the
wave function undergoes dispersion, which leads to revivals and super revivals
in the wave packet amplitudes. The classical motion and quantum revival
structure of this free time evolution allows for the possibility of
conveniently controlling the wave function at suitably chosen intervals of
time.

We propose using short optical pulses to couple coherently the Rydberg manifold
to a low-lying energy level in the atom. A transform-limited pulse that is
short compared to the classical Kepler period of the system interacts with all
the energy levels but couples only to the wave packet state that is nearest the
atomic core, a phenomenon familiar in the excitation and photoionization of
Rydberg states \cite{Wolde88}. The classical electron has the largest momentum,
and hence absorbs energy most efficiently, when near the core. During the
pulse, the wave packet state nearest the nucleus and the ground state form an
effectively two-level system, and undergo Rabi oscillations.

Coherent control of wave packets based on spatial localization has been studied
by Ahn et al. \cite{Ahn00}, who show that half-cycle pulses can be used to
extract binary phase information stored in a radial wave packet. The use of
optical pulses shorter than a Kepler period for quantum control of atomic wave
packets in the strong-field regime has been studied analytically by Araujo et
al. \cite{Araujo98}. A weak field analysis by Noel and Stroud \cite{Noel97}
showed that a uniform train of $d$ pulses within one Kepler period can be used
to create an arbitrary Rydberg state of $d$ levels whose amplitudes are related
by a discrete Fourier transform to the pulse amplitudes. We propose to use such
pulses for selective processing of quantum information in the wave packet
basis.

Consider $d$ Rydberg energy levels with angular momentum $l=1$ in a hydrogen or
alkali atom as a quantum information basis,
\begin{eqnarray}\label{energylevels}
    & \ket{j}_\nu = \ket{\nbar + j, 1, 0}, &
    \\* \nonumber
    & j = -d/2+1, -d/2+2, \ldots, d/2, &
\end{eqnarray}
where $\nbar$ is the mean value for the principal quantum number in the wave
packet superposition, and the subscript $\nu$ is used to denote a state in the
energy basis. We have assumed above that the number of levels $d$ is an even
number, but the arguments below are easily extended to odd $d$. These levels
can be excited simultaneously from the ground state by a strong linearly
polarized laser pulse with sufficient bandwidth to overlap all the levels. The
pulsed excitation creates a wave packet which evolves in time as
\begin{equation}\label{wavepacket}
    \ket{\psi(t)} = \sum_{n} \hs{0.2}
                     b_j \hs{0.2}\exp(-i\omega_j t) \hs{0.2}
                     \ket{j}_\nu,
\end{equation}
where $\hbar\omega_j$ is the energy of the $j$th Rydberg level and the
amplitudes $b_j$ are determined by the spectrum of the exciting pulse. If we
assume that the pulse spectrum, centered at $\omega_{0}$, is uniform over the
$d$ computational levels and falls off rapidly outside this manifold, we have a
uniform amplitude distribution, $b_j = 1/\sqrt{d}$, which is time independent
in the absence of external fields.

The radial wave packet was first studied by Parker and Stroud \cite{Parker86},
who showed that it corresponds to a radially localized shell of probability
distribution moving periodically between the classical turning points for short
times. Eventually, the uneven spacing in the energies of the levels in
Eq.~(\ref{energylevels}) leads to dispersions and revivals of the wave packet,
corresponding to the time-dependent phases in Eq.~(\ref{wavepacket})
interfering destructively or constructively. The rates at which the phases
evolve with respect to each other depend on the frequencies $\omega_j$ relative
to the average frequency $\omega_{0}$, which can be expanded in a Taylor series
in $j = n-\nbar$,
\begin{eqnarray}\label{Taylor}
    \omega_{j0} & = & \omega_j - \omega_{0}
        = - \frac{1}{2(\nbar+j)^2} + \frac{1}{2\nbar^2}
\nonumber \\*
    & = & 2\pi\left[\frac{j}{T_K} - \frac{j^2}{T_{\mathrm{rev}}} +
            \frac{j^3}{T_{\mathrm{sr}}} - \cdots \right],
\end{eqnarray}
%

%
\begin{figure}[htp]
\postscript{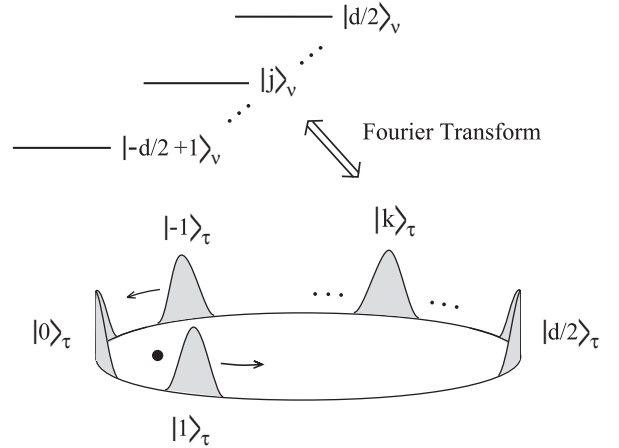}
\vs{1}\caption{Illustration of conjugate bases: Energy eigenstates
$\ket{j}_\nu$ related by a Fourier transform to wave packet states
$\ket{k}_\tau$ evenly distributed in time around a classical orbit. A radial
wave packet basis can be thought of as an ensemble of such orbits with
different orientations for the ellipses.}
\label{fig-WPbasis}\vs{1.9}
\end{figure}
%

\noindent where in atomic units, $T_K =2\pi\nbar^3$ is the classical Kepler
period for round trip motion, $T_{\mathrm{rev}}/2!=2\pi\nbar^4/3$ is when the
wave packet revives first, $T_{\mathrm{sr}}/3!=\pi\nbar^5/6$ is when the wave
packet super-revives first, etc. At $\nbar = 180$ for example, $T_K \simeq
0.89$ ns, $T_{\mathrm{rev}} \simeq 106$ ns, and $T_{\mathrm{sr}} \simeq 14$
$\mu$s. For large $\nbar$, the different time scales are well separated in
magnitude, and for a small number of levels $d$, the wave packet is more spread
out in space and disperses more slowly in time. For $d^2 \ll \nbar$, we see
from Eq.~(\ref{Taylor}) that the revival dynamics of the wave packet can be
neglected to good approximation during the first few Kepler periods, and
Eq.~(\ref{wavepacket}) becomes
\begin{eqnarray}\label{wavepacketkepler}
    \ket{\psi(t)}
    & \approx & \frac{1}{\sqrt{d}}
        \sum_{j} \exp\left(-i 2\pi j t/T_K\right)
        \ket{j}_\nu,
\nonumber \\*
    & = & \ket{\psi_K(t)}
\end{eqnarray}
which evolves periodically in time with period $T_K$.

The Fourier kernel in Eq.~(\ref{wavepacketkepler}) involves $j$ and $t$, and
can be thought of as underlying an uncertainty relation between energy and
time, with more levels leading to greater spatio-temporal localization of the
wave packet. Although the Rydberg energies are discrete, the time $t$ for the
wave packet varies continuously around the orbit. However, if we consider $d$
discrete times during any Kepler period,
\begin{eqnarray}\label{tm}
    & t_k = k (T_K/d), & \\*[0.2ex] \nonumber
    & k = -d/2+1, -d/2+2, \ldots, d/2, &
\end{eqnarray}
we can construct an alternate discrete wave packet basis $\ket{k}_\tau$ for the
system of energy levels in Eq.~(\ref{energylevels}) based on the quantum
Fourier transform (QFT),
\begin{eqnarray}\label{wavepacketbasis}
    \ket{k}_{\tau} & = & \ket{\psi_K(t_k)}
\nonumber \\*
    & = & \frac{1}{\sqrt{d}} \sum_{j}
            \exp\left(-i 2\pi j k /d \right) \hs{0.2}
            \ket{j}_\nu.
\end{eqnarray}
This is an orthogonal basis in time (denoted by the subscript $\tau$), and the
basis amplitudes evolve periodically with the classical period $T_K$, as
illustrated in Fig.~\ref{fig-WPbasis}. Thus $k=0$ and $k=d/2$ correspond to
wave packet states centered at the inner and outer turning points of the
classical orbit respectively. The sign of $k$ indicates the direction of
propagation of the wave packet. We can also define the basis for odd $d$,
except that in this case $k=\pm (d-1)/2$ correspond to wave packet states
centered near (but not exactly at) the outer turning point, having nonzero and
opposite momenta. The energy eigenstates can be recovered from $\ket{k}_\tau$
by an inverse QFT,
\begin{equation}\label{energylevelbasis}
    \ket{j}_\nu
    = \frac{1}{\sqrt{d}} \sum_{k}
        \exp\left(i 2\pi j k /d \right) \ket{k}_\tau.
\end{equation}

In general, any excited state \ket{\Psi(t)} of the $d$ Rydberg levels can be
expanded in either basis,
\begin{eqnarray}\label{Rydbergstate}
    \ket{\Psi(t)} & = &
    \sum_{j} b_j(t) \exp(-i\omega_j t)\hs{0.3} \ket{j}_\nu
\\*[0.2ex]
    & = & \exp(-i\omega_{0} t) \sum_{k} \tilde{b}_k(t) \hs{0.3} \ket{k}_\tau,
\end{eqnarray}
where we have removed the center frequency $\omega_{0}$ from the wave packet
amplitudes $\tilde{b}_k$. A classical discrete Fourier transform (DFT) relates
the two sets of amplitudes,
\begin{equation}\label{bn}
     \tilde{b}_k(t) = \frac{1}{\sqrt{d}}
                \sum_j  b_j(t) \exp(-i\omega_{j0} t) \hs{0.2}
                \exp\left(i2\pi j k/d\right).
\end{equation}
For free time evolution, $b_j(t) = b_j(0)$, and Eq.~(\ref{bn}) can be written
as
\begin{equation}\label{bmfree}
    \tilde{b}_k(t) = \sum_{k'} \tilde{b}_{k'}(0) \hs{0.4} u_{k'-k}(t),
\end{equation}
where $u_{k'-k}(t)$ is the DFT of the time evolution kernel,
\begin{equation}\label{um}
    u_{k'-k}(t) = \frac{1}{d} \sum_{j}
                    \exp\left[- i\omega_{j0} t - i2\pi j (k'-k)/d \right].
\end{equation}
In the Kepler regime of Eq.~(\ref{Taylor}), the levels are equally spaced,
$\omega_{j0} \simeq 2\pi j/T_K$, and $u_{k'-k}$ becomes a delta function in
$k'$, resulting in a cyclic permutation of the wave packet amplitudes in
Eq.~(\ref{bmfree}),
\begin{equation}\label{cyclicevol}
    \tilde{b}_k(nT_K/d) = \tilde{b}_{k-n}(0),
\end{equation}
where $n$ is an integer and $k-n$ is taken modulo $d$.
Equation~(\ref{cyclicevol}) amounts to a SHIFT gate in the wave packet basis,
the multivalued equivalent of the NOT gate for binary logic. More complex
unitary transforms occur for longer times, when the revival and superrevival
dynamics become relevant. Thus free time evolution alone becomes useful for
quantum information processing in the wave packet basis. This is a consequence
of the nonstationary character of the basis and the unequally spaced energy
levels.

We now turn to the control of the multilevel atomic basis using laser fields.
Universal multilevel transforms of the amplitudes $b_j$ can in principle be
achieved using narrow-band lasers to couple neighboring transitions among the
energy levels \cite{Muthukrishnan00}. However this requires multiple laser
frequencies tuned to the $d-1$ transitions. A simpler control scheme is
possible for the wave packet amplitudes $\tilde{b}_k$, as shown below. This
involves short broadband pulses and takes advantage of the cyclic evolution of
the wave packet states described above.

A laser pulse with a spectrum overlapping the $d$ Rydberg levels couples most
strongly to only that combination of the levels with a phase relation between
them that localizes the electron near the core. This corresponds to the wave
packet state $\ket{0}_\tau$, for which all the phases are equal since $k=0$.
Consider a pulsed field with a carrier frequency $\omega \approx \omega_{0}$
and pulse profile $f(t)$,
\begin{equation}\label{field}
    {\bf E}(t) = \mbox{\boldmath $\epsilon$} \hs{0.2} E_0 f(t)
                \exp(-i \omega t) + \mbox{c.c.},
\end{equation}
where $E_0$ and {\boldmath $\epsilon$} are the amplitude and polarization of
the field. Assume that the pulse couples the $d$ Rydberg levels to the ground
state \ket{g} in the atom. In the energy level basis for the Rydberg states,
the dipole interaction with the field yields the following coupled equations
for time evolution in the rotating wave approximation,
\begin{eqnarray}
\label{beneq}
    \dot{b}_g & = & \frac{i}{2} \hs{0.2} f(t)
        \sum_{j} \Omega_j \exp(-i\Delta_j t) \hs{0.2} b_j,
\\* \label{bneq}
    \dot{b}_j & = & \frac{i}{2} \hs{0.2} f(t) \hs{0.2}
        \Omega_j \exp(i\Delta_j t) \hs{0.2} b_g,
\end{eqnarray}
where $\Omega_j$ and $\Delta_j$ are the Rabi frequency and field detuning
defined as
\begin{eqnarray}\label{Rabin}
    \Omega_j & = & 2 \matrixelem{g}{\hat{\bf d}}{j}_\nu
                    \cdot \mbox{\boldmath $\epsilon$} \hs{0.2} E_0/\hbar,
\\ \label{Detuningn}
    \Delta_j & = & \omega_j - \omega,
\end{eqnarray}
and we have taken $\Omega_j$ to be real in Eqs.~(\ref{beneq}) and (\ref{bneq}).
Using the Fourier inverse of Eq.~(\ref{bn}), we rewrite the equation of motion
for $b_g$ in terms of the wave packet amplitudes,
\begin{equation}\label{bemeq}
    \dot{b}_g = \frac{i}{2} \hs{0.2} f(t) \exp(-i\Delta_{0} t)
        \sum_k \tilde{\Omega}_k \hs{0.2} \tilde{b}_k,
\end{equation}
where $\Delta_{0}$ is the mean detuning of the field, and $\tilde{\Omega}_k$ is
the DFT of the Rabi frequency $\Omega_j$,
\begin{equation}\label{Rabim}
    \tilde{\Omega}_k = \frac{1}{\sqrt{d}} \sum_{j}
                    \Omega_j \exp\left(-i2\pi j k /d\right).
\end{equation}
To find the equation of motion for the wave packet amplitude $\tilde{b}_k$, we
take the time derivative of Eq.~(\ref{bn}) and \linebreak

\begin{figure}[htp]
\postscript{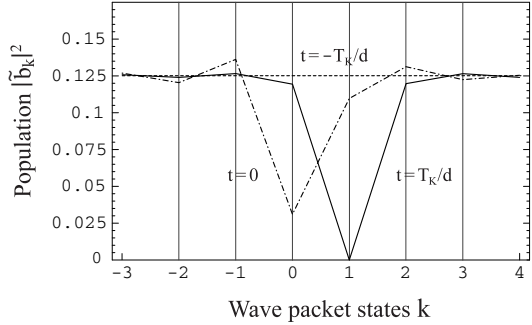}
\vs{1.3}\caption{Creation of a dark wave packet at $\nbar=180$ and $d=8$. The
spectral width (FWHM) of the $\pi$ pulse is approximately $1.3 d/T_K$. The wave
packet populations $|\tilde{b}_k|^2$ are shown before ($t=-T_K/d$), during
($t=0$) and after ($t=T_K/d$) the pulse.}
\label{fig-Pulse}\vs{1.3}
\end{figure}
%

\noindent use Eq.~(\ref{bneq}) for $\dot{b}_j$. Assuming that the pulse width
is short compared to a Kepler period, we can ignore the free atomic evolution
during the pulse that led to Eq.~(\ref{bmfree}). In the short pulse regime, we
are left with
\begin{equation}\label{bmeq}
     \dot{\tilde{b}}_k =
     \frac{i}{2} \hs{0.2} f(t) \hs{0.2}
            \tilde{\Omega}_k^* \exp(i\Delta_{0}t) \hs{0.2} b_g.
\end{equation}
Equations~(\ref{bemeq}) and (\ref{bmeq}) describe Rabi oscillations between the
ground state and the wave packet states, where the Rabi frequencies
$\tilde{\Omega}_k$ are Fourier transforms of their energy level counterparts
$\Omega_j$. For Rydberg levels, $\Omega_j$ varies slowly with $j$,
\begin{equation}\label{Rabinscaling}
    \Omega_j \sim (\nbar + j)^{-3/2},
\end{equation}
making only the dc component $\tilde{\Omega}_0$ significant for large $\nbar$.
For $\nbar=180$ and $d=8$, we find $\tilde{\Omega}_{\pm 1} \simeq 0.01 \mbox{
}\tilde{\Omega}_0$. Thus to good approximation, we can ignore the Rabi
frequencies $\tilde{\Omega}_{k \neq 0}$ in Eqs.~(\ref{bemeq}) and (\ref{bmeq}).
These equations reduce to
\begin{eqnarray}
\label{bemeqtwolevel}
    \dot{b}_g & \cong & \frac{i}{2} \hs{0.2} f(t) \tilde{\Omega}_0
        \exp(-i\Delta_{0} t) \hs{0.2} \tilde{b}_0,
\\* \label{bmeqtwolevel}
    \dot{\tilde{b}}_0 & \cong & \frac{i}{2} \hs{0.2} f(t) \tilde{\Omega}_0
        \exp(i\Delta_{0} t) \hs{0.2} b_g.
\end{eqnarray}
These describe two-level Rabi oscillations between the ground state and the
innermost wave packet state during the pulse, with a detuning $\Delta_{0} =
\omega_{0} - \omega$ between the peak of the pulse spectrum and the average
Rydberg frequency. The two-level approximation is valid only for short pulses.
To affect only the $k=0$ wave packet state, we need the pulse width $\tau_p$ to
be less than $T_K/d$, the time taken for this wave packet element to leave the
atomic core. However, the bandwidth of the pulse cannot be much larger than the
atomic frequencies spanning the $d$-level basis, to avoid coupling into levels
outside this basis. Since the Rydberg frequency spacing is approximately
$1/T_K$, this implies a bandwidth $\Delta\omega /2\pi \simeq d/T_K$. To
summarize, we require that

\begin{figure}[htp]
\postscript{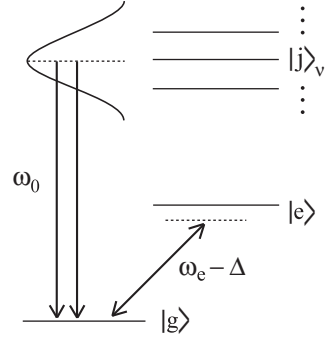}
\vs{1}\caption{Wave packet control for quantum information processing. The
energy level basis is $\ket{j}_\nu = \ket{\nbar+j,1,0}$. Broadband $\pi$ pulses
de-excite two wave packet amplitudes in the conjugate wave packet basis. The
ground levels $g$ and $e$ are coupled by a laser with frequency $\omega \approx
\omega_e$.}
\label{fig-WPlogic}\vs{2.2}
\end{figure}
%

\begin{equation}\label{pulsecons}
    \tau_p < \frac{T_K}{d} \simeq \frac{2\pi}{\Delta\omega},
\end{equation}
or that the pulse be transform-limited. For $\nbar = 180$ and $d=8$ levels, we
use a Gaussian pulse of width $\tau_p = 0.5 \ln 2$ $T_K/d \simeq 110$ ps (FWHM
amplitude), corresponding to a spectral width $\Delta\omega/2\pi = (8/2\pi)$
$d/T_K$ for the electric field. Figure~\ref{fig-Pulse} shows an exact numerical
solution for the wave packet amplitudes before, during and after the pulse.
Assuming that the population is uniformly distributed initially, we find that a
$\pi$ pulse transfers nearly all the population in the $k=0$ state to the
ground state, creating a ``dark'' wave packet in the Rydberg basis that evolves
cyclically in the Kepler evolution. About $5\%$ of the population is lost from
neighboring states $k=\pm 1$ during the pulse. Thus a coherent coupling of
individual wave packet states with the ground state can be modeled as a
two-level system to good approximation.

Since the wave packet basis is nonstationary in time, the Rydberg wave function
undergoes dispersion and revivals as the energy level phases evolve in the \Sch
picture. This leads to aperiodicity in the free time evolution of the wave
packet amplitudes $\tilde{b}_k$, going beyond the cyclic regime of
Eq.~(\ref{cyclicevol}). At $\nbar = 180$ and $d=8$ for example, we find that
the population in a wave packet state decays by about $5\%$ after a Kepler
period, but recovers to within $2$-$4\%$ at a revival or super-revival. This
scales as $\nbar^{-2}$, making dispersion losses less significant for larger
$\nbar$. By appropriate timing of successive laser pulses, we can thus take
advantage of the revival structure of the atomic energy spectrum when
addressing the wave packet amplitudes in the basis.

An arbitrary unitary transform of the $d$ wave packet states can be decomposed
into a product of ${\cal O}[d^2]$ two-dimensional transforms \cite{Ekert96},
each of which can be implemented sequentially on two de-excited wave packet
amplitudes at a time, as shown in Fig.~\ref{fig-WPlogic}. Two low lying energy
levels $g$ and $e$ temporarily store the wave packet amplitudes in the form
$\tilde{b}_k \ket{g} + \tilde{b}_{k'} \ket{e}$. Phase shifts or rotations of
this state are accomplished by coupling the two levels with a pulse of
appropriate length, Rabi frequency and detuning. The transformed state is then
restored to the Rydberg basis. Thus multilevel transforms in the atom are
possible with a timed sequence of laser pulses, without the need for multiple
lasers to address the different Rydberg transitions. Moreover the fidelity of
the gate operations compares favorably with that in an equivalent system of
entangled two-level atoms \cite{Sackett00}.

For efficiently scalable quantum computing, we require conditional transforms
on two entangled multilevel atoms. The method used in Ref. \cite{Sackett00} is
based on the linear ion trap scheme for binary quantum logic \cite{Cirac95}.
This scheme provides a means to evolve one two-level atom in the trap array
conditional on the state of another two-level atom. Since the wave packet
approach allows each basis amplitude in the atom to be addressed sequentially
through low-lying energy levels, conditional transforms on other atoms can be
carried out using the same methods as in the binary scheme. That is,
conditional on a given wave packet state (vis \`{a} vis energy level) in the
control atom, a $d$-dimensional transform is performed on the wave packet basis
of the target atom using the methods outlined above.

The time taken to couple two atoms in this approach can be on the order of the
atomic Kepler period, which permits a fast non-adiabatic coupling scheme. Also,
the mean energy and number of Rydberg levels used depend ultimately on the
ionization threshold imposed by external fields in the environment. We
anticipate that neutral atom schemes that allow fast coupling times, as have
been suggested recently \cite{Jaksch00}, will provide a useful setting for
multilevel quantum information applications.


This work was supported by the Army Research Office through the MURI Center for
Quantum Information.

\vs{-0.5}
\makeatletter


\end{document}